\documentclass[conference]{IEEEtran}
\IEEEoverridecommandlockouts
\usepackage{cite}
\usepackage{amsmath,amssymb,amsfonts}
\usepackage{algorithmic}
\usepackage{graphicx}
\usepackage{textcomp}
\usepackage{physics}
\usepackage{xcolor}
\usepackage{soul}
\usepackage{hyperref}
\usepackage[normalem]{ulem}
\usepackage{url}

\usepackage{cleveref}

\def\BibTeX{{\rm B\kern-.05em{\sc i\kern-.025em b}\kern-.08em
    T\kern-.1667em\lower.7ex\hbox{E}\kern-.125emX}}

\usepackage{lipsum}

\usepackage{commondefs}

\DeclareMathOperator{\calV}{\mathcal{V}}
\DeclareMathOperator{\calM}{\mathcal{M}}
\newcommand{\vecX}{\mathbf{X}}
    
\begin{document}

\title{Quantum Hypergraph Partitioning}

\author{
\IEEEauthorblockN{
Cameron Ibrahim\IEEEauthorrefmark{1},
Bao G. Bach\IEEEauthorrefmark{1}\IEEEauthorrefmark{2},
Jad Salem\IEEEauthorrefmark{3},
Reuben Tate\IEEEauthorrefmark{4},
Kien X. Nguyen\IEEEauthorrefmark{1},
Stephan Eidenbenz\IEEEauthorrefmark{4},
Ilya Safro\IEEEauthorrefmark{1}
}
\IEEEauthorblockA{\IEEEauthorrefmark{1}\textit{Computer and Information Sciences, University of Delaware}, Newark, DE, USA}
\IEEEauthorblockA{\IEEEauthorrefmark{2}\textit{Quantum Science and Engineering, University of Delaware}, Newark, DE, USA}
\IEEEauthorblockA{\IEEEauthorrefmark{3}\textit{United States Naval Academy}, Annapolis, MD, USA}
\IEEEauthorblockA{\IEEEauthorrefmark{4}\textit{Los Alamos National Laboratory}, Los Alamos, NM, USA}
\IEEEauthorblockA{\{cibrahim@udel.edu, baobach@udel.edu,  isafro@udel.edu\}}
}

\maketitle

\begin{abstract}
Quantum optimization algorithms are inherently probabilistic, yet they are most often used to search for a single high-quality solution. In this paper, we instead study hypergraph partitioning problems in which the desired output is itself a probability distribution over partitions. We introduce a distributional perspective on hypergraph partitioning motivated by maximin and minimax objectives such as Fair Cut Cover, and we show how these objectives align naturally with the measurement distribution produced by QAOA. To motivate the formulation, we introduce a workforce-scheduling-inspired toy problem, the Greatest Expected Imbalance problem, in which the goal is to minimize the worst expected imbalance across hyperedges. We then develop QAOA-based quantum solvers that represent distributional solutions natively through quantum states, together with quadratic hypergraph objectives suitable for standard and multi-objective QAOA. These formulations connect balanced hypergraph partitioning, polarized community discovery, and distributional fairness under a unified quantum optimization framework. For comparison, we provide optimal polynomial-time classical approximation algorithms based on semidefinite programming and hyperplane rounding. Experiments on real-world and synthetic hypergraphs demonstrate that low-depth multi-angle QAOA can outperform these classical approximation baselines on the proposed
objectives, highlighting the potential of quantum algorithms for optimization problems where the solution is a distribution rather than a single partition.\\

%Hypergraph Partitioning is a ubiquitous problem in scientific computing concerned with partitioning a set of nodes into two or more distinct sets while either minimizing or maximizing the number of connections between them. In this paper, we introduce a distributional perspective on Hypergraph Partitioning based on distributional maximin optimization problems such as Fair Cut Cover. To motivate our formulation, we take inspiration from Workforce Scheduling to introduce a toy problem called the Minimum Expected Imbalance problem. To solve this problem, we provide a quantum framework based on QAOA, utilizing the probabilistic nature of quantum computing to represent the distributional solutions natively. Furthermore, we introduce a quadratic objective for Hypergraph Partitioning, which is amenable to standard QAOA solvers, which we use to define a multi-objective formulation motivated by Balanced Hypergraph Partitioning and Polarized Community Discovery. For comparison, we provide the best possible polynomial-time algorithm for each of the defined problems. We then evaluate our method on a collection of hypergraphs derived from real-world examples, as well as synthetic examples chosen to demonstrate the gap between the quantum solvers.\\

\noindent{\textbf Reproducibility}: source code and data are available at \url{https://github.com/cameton/QuantumHypergraphPartitioning}\\

\end{abstract}


\section{Introduction}\label{sec:intro}

Most research on quantum algorithms for optimization \cite{abbas2024challenges} has focused on finding a single high-quality solution. But why should naturally probabilistic quantum computers be forced to solve deterministic optimization problems? Most quantum optimization workflows treat measurement randomness as a nuisance, repeatedly sampling a circuit only to extract one best solution. In this paper, we take the opposite view: when the application itself calls for a distribution over partitions, the probabilistic output of a quantum algorithm becomes the solution object rather than a byproduct.

Hypergraphs are an abstract mathematical structure used to model relationships between groups of individuals, generalizing the pairwise relationships more commonly modeled by graphs. Hypergraph Partitioning is a ubiquitous problem in scientific computing with applications in fields such as quantum circuit simulation \cite{cotengra}, VLSI design \cite{cong2013multilevel}, social network analysis \cite{hypergraph_social_analysis1, hypergraph_social_analysis2}, machine learning \cite{gao2020hypergraph,sybrandt2020fobe,zhou2007learning}, and operations research \cite{transitrouting}. Hypergraph Partitioning problems are strongly related to the separability of a given hypergraph structure, and oftentimes serve as an important component to a number of other algorithms as a preprocessing step for identifying local regions in a given input \cite{scalable_graph}.

A classical formulation of Hypergraph Partitioning generally aims to find a single assignment of the nodes onto two disjoint sets in order to minimize the size of the cut, i.e., the number of hyperedges which contain at least two vertices assigned to different sets, while maintaining some balance constraint on the size of the two\footnote{While in general, the Hypergraph Partitioning is defined on $k$ disjoint partitions, in our paper $k=2$, which still keeps the problem NP-hard.} partitions \cite{kahypar,sybrandt2020hypergraph,shaydulin2019algdist}. 
If instead the aim is to maximize the size of the cut, this problem is known as the Max Set Splitting problem\cite{quantumsetsplitting}. 
Hypergraph Partitioning serves as a generalization of another common problem: Graph Partitioning \cite{buluc2016recent}. The most common approach to both problems assigns an integer label to each vertex of the input and scores each edge/hyperedge based on whether there is a difference between labels of any two vertices in that edge/hyperedge \cite{recentadvances}. 
% However, while this is a quadratic objective for graph partitioning, the same is not true for hypergraphs.\BB{Require citation for this statement?} 

Distributional optimization is a branch of optimization concerned with finding a distribution over solutions to a problem which, in expectation, achieves a desired outcome \cite{distributional}. These frameworks are often challenging to handle classically, potentially requiring many solutions for the target problem to be found and maintained in memory \cite{bach2026learning}.  However, there can be significant benefits; for example, by sampling a variety of solutions over a given period of time, one can achieve a lower expected cost overall or achieve a fairer outcome on average \cite{maximin_fair}. Furthermore, these problems are naturally suited for quantum computing, as the probabilistic nature of quantum computers means these distributions over solutions can be represented natively by algorithms such as QAOA \cite{qaoaMaxCut2018}.

% Quadratic optimization via quantum computing has received a great amount of attention by the quantum community, with proposed quantum solutions for a wide variety of quadratic problems including graph partitioning \cite{ushijima2021multilevel}. One of the most notable of these strategies is the quantum-classical Quantum Approximate Optimization Algorithm (QAOA), which constructs a parameterized quantum ansatz based on an input graph which is then optimized using classical solvers. Recently, the QAOA algorithm was extended to a multi objective setting via Multi-Objective QAOA (mo-QAOA), which simultaneously optimizes a given collection of quadratic problems \cite{moqaoa}. In order to leverage this wealth of existing research on optimizing quadratic problems using quantum methods, it would be very valuable to have a quadratic form of hypergraph partitioning.

%\is{There is a disconnect between intro to hypergraphs and our contributions. There should be some story about distributional variations. Something similar to what we have in the previous paper.}
%\CI{How's that? See above}

\textbf{Our Contribution:} In this paper, we provide a distributional perspective on the Hypergraph Partitioning problem, with the aim of utilizing the probabilistic nature of quantum computing to natively represent distributional solutions to maximin and minimax optimization problems. In particular, we provide the following contributions: 
\begin{itemize}
\item 
We formalize three different variations of the distributional Hypergraph Partitioning problem, which we believe are well-suited to a quantum approach.

With the first, we introduce a distributional formulation of Hypergraph Partitioning, which we call the Greatest Expected Imbalance problem, which aims to find a distribution over partitions that minimizes the greatest expected imbalance across hyperedges in the hypergraph, a quadratic measure of the separation of vertices within a given edge.  The second formulation examines the total imbalance of edges in the hypergraph, which produces a problem that directly generalizes the standard graph partitioning formulation. Finally, we extend this problem to create a multi-objective formulation of the Hypergraph Partitioning problem in order to generalize the concept of Balanced Hypergraph Partitioning. 

\item We introduce the hybrid quantum-classical approach used to solve these three problems based on QAOA. Furthermore, we provide classical approximation algorithms for each problem based on semidefinite programming and hyperplane rounding. 

\item We analyze the provided algorithms, including proofs relating the introduced problems to previous formulations of Hypergraph Partitioning, and explain how these relationships show that the provided approximation algorithms cannot be improved upon as long as the Unique Games Conjecture holds.

\item Finally, we support this analysis with empirical experiments on inputs generated from real-world hypergraphs, as well as synthetic graphs that have been chosen to demonstrate the gap between our quantum approach and classical approximations.
\end{itemize}  

\begin{figure*}
    \centering
    \includegraphics[width=1.0\linewidth]{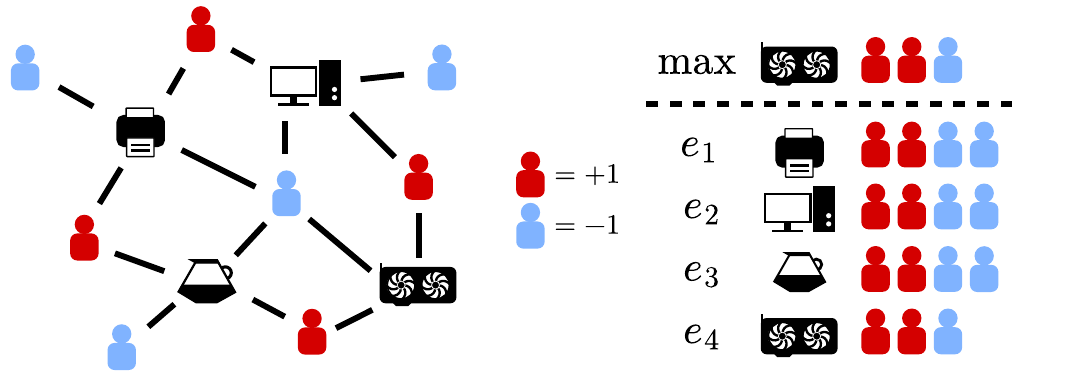}
    \caption{An example of the hyperedge imbalance arising from a single shift assignment in a Workforce Scheduling context. A hypergraph (left) contains vertices representing employees (red and blue figures) and hyperedges representing resources required by those employees during their shift (black). The aim is to assign each employee \(v\) a shift \(x_v \in \{\pm 1\}\) corresponding to the colors red and blue, so to minimize the greatest imbalance in employees across all resources. We measure imbalance as the square of the average value for employees who require a given resource. We assume that each resource \(e\) is equally important and that demand for each resource is uniform, so \(w_e = 1\) and \(P_{ve} = 1/\abs{e}\) for all \(v,e\). As such, \(w_e(p_e^*x)^2 = \abs{\sum x_v}^2/|e|^2\).  For the provided example, the most imbalanced resource is the GPU, which has 2 people in shift \(+1\) and one in shift \(-1\), yielding a hyperedge imbalance of \(1/9\). } 
    \label{fig:main_fig}
\end{figure*}

\section{Background}\label{sec:background}

In this section, we will introduce the necessary background and notation required to understand the rest of the paper.

% TODO add definition of frobenius norm
\subsection{Basic Definitions \& Notation}

We denote the vector of all ones as \(\One\) and the vector of all zeros as \(\Zero.\) The transpose of a real matrix \(M\) is a matrix \(M^*\) such that \((M^*)_{ij} = M_{ji}\).  \(M\) is symmetric if \(M^* = M.\) \(M\) is positive semi-definite if \(x^*Mx \geq 0\) for all \(x \in \Real^n.\) If \(M\) is symmetric and positive semi-definite (spsd), then so is \(A^*MA\) for any real matrix \(A.\) The Frobenius inner product between real matrices \(A\) and \(B\) is given by $\langle A, B \rangle_F  = \Tr{A^{*} B}$.

In this paper, we focus primarily on real vector valued random variables \(\vecX\) taking values on \(S=\{\pm 1\}^n\), which we'll refer to as random spin vectors.  These variables implicitly model a discrete distribution over spin variables, which may be recovered by evaluating \(\Prob[\vecX = x]\) for some \(x \in \{\pm 1 \}^n\). For some function \(f\), the expectation of \(f\) on \(\vecX\) is given by 
\[
\Exp[f(\vecX)] \coloneqq \sum_{s \in S} \Prob[\vecX = s]f(s).
\] 
The autocorrelation of \(\mathbf X\) is defined as \(\Exp[\vecX\vecX^*]\), which is symmetric and positive semi-definite. Autocorrelation is traditionally denoted \(R_{\mathbf{XX}}\) in probability literature, but we will use \(Q_{\mathbf{X}}\) in this paper to avoid confusion with the rotation X gate. For any random spin vector \(\vecX\), \(\diag(Q_\vecX) = \One.\)

The Kronecker delta function
\(\delta(x)\) returns 1 if the input is nonzero, and 0 otherwise.
 
A multi-objective optimization problem consists of a tuple of objective functions \(f(x)  = \left(f_i(x)\right)_{i \in \Index}\) for some index set \(\Index.\) The goal is to find a Pareto front for these functions, that is a collection of points \(x\) for which there does not exist a \(y\) which is strictly better on every objective. When \(\abs{\Index} = 2\), one can estimate the Pareto front by solving \(\max_x \alpha f_1(x) + (1-\alpha)f_2(x)\) for a series of \(\alpha \in [0, 1].\)

\subsection{Hypergraph Partitioning}
\label{sec:hypergraph_partitioning}
A hypergraph \(H = (V, E)\) is a set of vertices \(V\) along with a set of hyperedges \(E \subseteq \Pow(V) \setminus \{\emptyset\}.\) If every edge \(e \in E\) has \(\lvert e \rvert = 2,\) then \(H\) is known as a graph. A hypergraph will often have a set of edge weights \(w_e\) for each \(e \in E.\)

For a given hypergraph $H = (V, E)$, we define its \emph{clique expansion} to be a graph $G = (V', E')$ where $V' = V$ and $E = \{ (u,v) \in {V \choose 2} : \exists e \in H \text{\ s.t. \ } \{u,v\} \subseteq e\}$, i.e., edges are formed in $G$ by transforming each hyperedge in $H$ into a clique connecting all of that edge's vertices.

A hypergraph \(H\) is representable as an incidence matrix \(B_{ve} = \mathbbm{1}[v \in e]\) where $\mathbbm{1}[\cdot]$ is the indicator function. A graph has an additional representation as an adjacency matrix \(A_{uv} = w_{uv}\mathbbm{1}[uv \in E],\) as well as the graph Laplacian \(L = \diag(A\One) - A.\)
As a quadratic form, this is equivalent to \(x^*Lx = \sum_{uv \in E} w_{uv}(x_u - x_v)^2.\)

Recall \(x\) is a string that denotes a set of vertices with $x_u = 1$ if vertex $u$ is in that set. We say \(x\) cuts \(e\) if \(\exists u,v \in e : x_u \neq x_v.\)
For 2 partitions, the standard formulation of Balanced Hypergraph Partitioning\cite{kahypar} can be written as
\begin{align*}
    \min_{x \in \{\pm 1\}^{\lvert V \rvert}} \sum_{e\in E} w_e \Ind[x \text{ cuts } e]
    \text{ st } (\mathbf{1}^* x )^2 \leq \beta.
\end{align*}
If \(H\) is a graph, this is equivalent to the quadratic programming problem
\begin{align*}
    \min_{x \in \{\pm 1\}^{\lvert V \rvert}} \frac{1}{4}x^*Lx \text{ st } (\mathbf{1}^* x )^2 \leq \beta.
\end{align*}

Alternatively, one can attempt to maximize the number of edges in the cut, yielding the unconstrained problems
\begin{align*}
     \max_{x \in \{\pm 1\}^{\lvert V \rvert}} &\sum_{e\in E} w_e \Ind[x \text{ cuts } e]& \max_{x \in \{\pm 1\}^{\lvert V \rvert}} & \frac{1}{4} x^*Lx,
\end{align*}
which are known as the Set Splitting problem and the MaxCut problem, respectively.
Balanced Hypergraph Partitioning, the Set Splitting problem, and Max Cut are all examples of NP-Hard and APX-Hard optimization problems \cite{bach2026learning}.

\subsection{Distributional Optimization} % TODO expand this
A probability distribution on a finite set \(\Omega\) is a nonnegative real vector \(q \in \Real^{\lvert \Omega\rvert}_{\geq 0}\) with \(\lVert q \rVert_1 = 1.\) 
Let \(f_s\) be a nonnegative function for each \(s\) in a population \(S\). We utilize a minimax variation of the framework for expected maximin fairness outlined in \cite{maximin_fair}
\begin{align*}
    \min_{\vecX}\max_{s \in S} &\ \Exp[f_s(\vecX)].
\end{align*}
These can be interpreted as finding a distribution over solutions such that the Greatest Expected Cost / Least Expected reward is minimized/maximized in the population.

One example of a distributional optimization problem is the Fair Cut Cover problem, where the objective is to find a distribution over possible partitions which maximizes the lowest probability that any edge is cut. Specifically, for a graph \(G\), you have \(S = E\) with \(f_{uv}(x) = \frac{1}{4}(x_u - x_v)^2.\) Fair Cut Cover is a challenging problem in general, as it is both NP-Hard and APX-Hard \cite{bach2026learning}.

% END Distributional Optimization

\subsection{Approximation via Semidefinite Programming}

Assuming the Unique Games Conjecture, the optimal polynomial-time approximation algorithms for Max Cut and Fair Cut Cover are both achieved via Semidefinite Programming and hyperplane rounding \cite{khot2007optimal, bach2026learning}.

Originally introduced by Goemans and Williamson, this approach relaxes a quadratic optimization problem into a semidefinite programming problem, a convex optimization problem where the solution comes in the form of an spsd matrix and can be solved in polynomial time\cite{goemans1995improved}. For example, the semidefinite programming relaxation of Max Cut is given by 
\begin{align*}
        \max_{A \text{ spsd}} \frac{1}{4}\frob{L, A} \text{ st } \diag(A) = \One.
\end{align*}

Because all spsd matrices can be decomposed as\(A = BB^*\), one can utilize a process called hyperplane rounding to generate a random spin vector \(\vecX.\)
To do so, we define a vector valued normal random variable \(\mathbf{Y} \sim \mathrm{Normal}(0, I)\) and define \(\mathbf{X} = \mathrm{sign}(B\mathbf{Y}).\)
For Max Cut, this process essentially embeds the vertices of your graph as the rows of \(B,\) then samples a random hyperplane to partition these points. The probability that an edge \(uv\) is cut by this plane is given by
\begin{align*}
    \frac{1}{4}\Exp\left[\abs{\vecX_u - \vecX_v}^2\right] =  \frac{1}{\pi}\arccos{(A_{uv})},
\end{align*}
meaning that the expected Max Cut value is given by 
\begin{align*}
    \frac{1}{4}\Exp[\vecX^*L\vecX] = \frac{1}{\pi}\sum_{uv \in E} \arccos(A_{uv}).
\end{align*}
Goemans and Williamson proved that this procedure achieves an expected approximation ratio of
\[
\alpha_{\mathrm{GW}} \approx 0.87856
\]
with respect to the optimal cut value \cite{goemans1995improved}.

\subsection{Quantum Approximate Optimization Algorithm}
The Quantum Approximate Optimization algorithm (QAOA) \cite{farhi2014quantum} is a hybrid quantum algorithm designed for combinatorial optimization problems that can be encoded as a quadratic form. Given this, QAOA with $p$ layers alternatingly apply unitaries drawn from two Hamiltonian families, cost unitary $U_{C}(\gamma) = e^{-i\gamma \hat{H}_C}$ and mixing unitary $U_{M}(\beta) = e^{-i\beta \hat{H}_M}$ parametrized by $\gamma = \{\gamma_i\}$ and $\beta = \{\beta_i\}$, $1\leq i \leq p$, respectively. The Hamiltonian $\hat{H}_C$ is a cost Hamiltonian where the information of the given quadratic optimization problem is embedded, while Hamiltonian $\hat{H}_M$ is a fixed mixing Hamiltonian. Using $\hat{H}_C$ as the observable, QAOA prepares the quantum state expressed in  Equation (\ref{eq: QAOA_state}):
\begin{equation}
    \label{eq: QAOA_state}
    \ket{\gamma,\beta} = U_{M}(\beta_p)U_{C}(\gamma_p)\dots U_{M}(\beta_1)U_{C}(\gamma_1)\ket{+}^{\otimes n},
\end{equation}
and performs optimization over the parameters $\gamma$ and $\beta$ with respect to the expectation value $\langle \hat{H}_C \rangle = \bra{\gamma,\beta}\hat{H}_C\ket{\gamma,\beta}$ to find the minimum.

\subsection{Related Work}
This distributional viewpoint of the hypergraph problem is especially relevant in quantum computation, where the output of an algorithm is inherently probabilistic. Rather than treating this probabilistic output as an obstacle that must be post-processed into one good solution, we treat it as the optimization target itself. Prior work in quantum generative modeling has explored whether parameterized quantum circuits can learn or engineer useful probability distributions \cite{liu2018differentiable, gili2023quantum}. Following the foundation in \cite{bach2026learning, salem2024expected}, we consider this hypergraph problem in a different setting by studying
combinatorial optimization problems in which the goal is to learn a distribution that maximizes a worst-case criterion over the underlying discrete structure. 

The balanced constraints in the hypergraph partitioning can be formulated as a multi-objective problem. \cite{kotil2025quantum} presents Quantum Approximation Multi-Objective Optimization, which performs multi-objective optimization over different MaxCut instances. Our formulation is inspired by this setting, in which we formulate the balance constraints over the complete graph.

\section{Groundwork}\label{sec:groundwork}

In this section, we will introduce a distributional hypergraph partitioning problem motivated by analogy to Workforce Scheduling\cite{castillo2012survey}. To do so, we introduce notions of the imbalance and variance of a hyperedge with respect to a vector \(x.\) 
Furthermore, we introduce a multi-objective formulation of hypergraph partitioning motivated by  balanced graph partitioning and Polarized Community Discovery\cite{polarization}. In order to do this, we introduce a quadratic hypergraph partitioning objective based on the total variance of all hyperedges.

\subsection{Distributional Hypergraph Partitioning}

Say you have a population of employees who need to be assigned to one of two shifts throughout the day. Each employee needs access to a set of resources in order to do their job. In order to reduce competition for a given resource, the employees who require that resource should be balanced between the two shifts. 

To each resource, we assign a set \(e\) containing the employees who will require that resource during their shift. The proportional amount that employee \(v\) requires resource \(e\) is given by \(P_{ve},\) where each column of \(P\) sums to one. We denote the column of \(P\) corresponding to a resource \(e\) as \(p_e.\) Let \(x_v \in \{\pm 1\}\) indicate the shift assigned to employee \(v.\) The imbalance in employees who require a resource \(e\) can then be measured as \(( p_e^* x)^2\) and the relative impact of that imbalance can be indicated by a weight \(w_e\), which yields the minimization problem: 
\begin{align*}
    \min_x \max_e w_e(p_e^*x)^2.
\end{align*}
However, we may consider varying employee shifts over the course of some period of time, meaning that although a resource may be imbalanced for one shift assignment, it might not be imbalanced for another. If we determine which shift to use at a given point in time using a distribution of possible shift assignments, we could instead consider minimizing the Greatest Expected Imbalance of the hypergraph. Letting \(\vecX\) be a random spin vector representing a random shift assignment and \(\calM_e = p_ep_e^*\), the Greatest Expected Imbalance problem is defined as
\begin{align*}
    \min_\vecX \max_e w_e\Exp[\vecX^*(p_ep_e^*)\vecX] = \min_\vecX \max_e w_e\frob{\calM_e, Q_\vecX},
\end{align*}
where \(Q_\vecX = \Exp[\vecX\vecX^*]\) is the autocorrelation matrix of \(\vecX.\) A visualization of this problem is provided in \cref{fig:main_fig}. From here, we can give a formal definition of this problem in terms of hypergraphs.

We introduce a weighted variant of the incidence matrix of a hypergraph, which we refer to as a stochastic incidence matrix. We define a stochastic incidence matrix for \(H\) as a nonnegative \(\lvert V\rvert\times \lvert E\rvert\) matrix \(P\) where each column sums to one and \(P_{ve} \neq 0\) if and only if \(v \in e.\) For convenience, we define \(p_e\) as the column of \(P\) corresponding to edge \(e\). 

Given a vector \(x,\) we can define a random variable \(X_e\) for each hyperedge \(e\) which takes the value \(x_v\) with probability \(\Prob[X_e = x_v] = P_{ve}.\) Then the expectation of \(\Exp[X_e] = p_e^*x\), in essence the average of values the vector \(x\) has on the vertices of \(e\).

We define the imbalance of an edge \(e\) for a vector \(x\) as the squared expectation of \(X_e\), \(\Exp[X_e]^2 = ( p_e^*x)^2 = x^*\calM_e x\). The variance of \(X_e\) is given by
\begin{align*}
    \Var[X_e] = x^*\calV_e x = \sum_{v}P_{ve}(x_v - p_e^*x)^2
\end{align*}
where \(\calV_e \coloneqq \diag(p_e) - \calM_e\). We refer to \(\Exp[X_e]^2\) and \(\Var[X_e]\) as the imbalance and variance of \(e\) with respect to \(x,\) and to \(\calM_e\) and \(\calV_e\) as the imbalance and variance matrices of \(e.\)

This allows us to define a maximin variant of the Greatest Expected Imbalance problem, which we refer to as the Least Expected Variance problem
\begin{align*}
    \max_{\vecX} \min_e w_e\Exp[\vecX^*\calV_e\vecX] = \max_{\vecX} \min_e w_e\frob{\calV_e, Q_\vecX}.
\end{align*}
Because \(x^*\diag(p_e)x = 1\) for any \(x \in \{\pm 1\}^{\lvert V\rvert}\), \(x^*\calV_e x = 1 - x^*\calM_e x\) for all \(e \in E\). As such, if each hyperedge \(e\) has weight \(w_e = 1\), then the Greatest Expected Imbalance and Least Expected Variance problems are equivalent.

\subsection{Multi-objective Hypergraph Partitioning}

\begin{figure}
    \centering
    \includegraphics[width=1.0\linewidth]{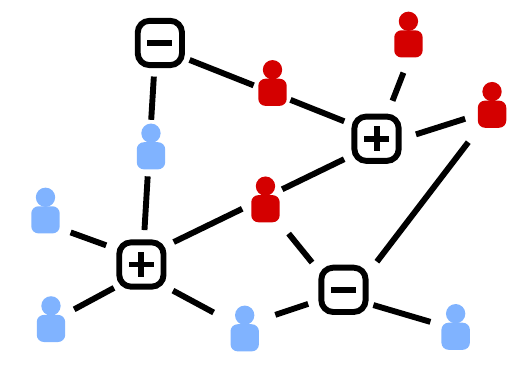}
    \caption{An example of a signed hypergraph which is amenable to the Multiobjective Hypergraph Partitioning problem. Each \(+\) hyperedge indicates individuals in that hyperedge should be placed together, while a \(-\) hyperedge indicates that these individuals should be placed separately. This signed hypergraph can be decomposed into a hypergraph consisting of all \(+\) hyperedges and all \(-\) hyperedges; simultaneously minimizing the variance of the first hypergraph while mazimizing the imbalance of the second achieves the desired affect.}
    \label{fig:polarization}
\end{figure}

In this section, we take inspiration from standard formulations of balanced graph partitioning and Polarized Community Discovery to develop a multi-objective approach to hypergraph partitioning. To do so, we introduce a quadratic hypergraph partitioning objective which is amenable to standard QAOA approaches and can easily be rescaled to better explore the Pareto front.

In Polarized Community Detection literature, the edges of a graph are allowed to take negative weights; in this way, a cut minimization problem becomes an attempt to cut negatively signed edges and to avoid cutting positively signed edges \cite{polarization}. In this way, the partitioning problem simultaneously tries to group people who like each other while separating individuals who disagree. 
If we were to extend this notion to our concepts of hypergraph imbalance and variance, we could consider trying to maximize the imbalance of some hyperedges (grouping people together), while maximizing the variance of others (moving people further apart). An example of this can be seen in \cref{fig:polarization}. One can also interpret this as a pair of individuals having an attractive force between them if they appear together in a hyperedge in one hypergraph, and a repelling force between them if they appear together in a hyperedge in the other. If the two individuals appear together in both hypergraphs, we can consider what sort of a net force occurs between them. 

We can find a similar dynamic in balanced graph partitioning.
Recall that the quadratic programming formulation of balanced graph partitioning is given by
\begin{align*}
    \min_{x \in \{\pm 1\}^{\lvert V \rvert}} \frac{1}{2}x^*Lx \text{ st }( \mathbf{1}^* x )^2 \leq \beta.
\end{align*}
Constrained optimization in a quantum setting is often challenging, making a penalty approach to optimization an appealing option \cite{abbas2024challenges, chen2025slack}. The penalty formulation of balanced graph partitioning is given by 
\begin{align*}
    \min_{x} \frac{1}{2} x^*(L+ \lambda \One\One^*) x \text{ st } x \in \{\pm 1\}
\end{align*}
\cite{budgetconstrained}. This is again a problem of simultaneously moving people closer together (the Laplacian term) while also moving them further apart (the balance term).

To further develop this notion of simultaneously moving objects closer together and further apart, we introduce quadratic formulations of hypergraph partitioning which extend notions of hyperedge imbalance and variance to the hypergraph as a whole. 
Since we have a pair of problems concerned with looking at the greatest imbalance and least variance over the set of hyperedges, it is natural to consider the sum of all imbalances or the sum of all variances. This yields a quadratic form, which serves as an effective quadratic objective for hypergraph partitioning.

We define the imbalance of \(H\) given \(x\) as the weighted sum of imbalances for each hyperedge given \(x.\)   Likewise, the  variance of \(G\) given \(x\) is the weighted sum of variances for each hyperedge given \(x\)
\begin{align*}
    x^*\calM x &= \sum_e w_e(p_e^*x)^2, & x^*\calV x &= \sum_{v,e} w_eP_{ve}( x_v - p_e^*x)^2
\end{align*}
Where \(\calM \coloneqq P\diag(w)P^*\) and \(\calV \coloneqq  \diag(Pw) - \calM\). We refer to \(\calM\) and \(\calV\) as the Imbalance and Variance Matrices of \(H.\)

With these quadratic forms, we can then define the Total Variance and Total Imbalance problems,
\begin{align*}
    \min_\vecX &\frob{\calM, Q_\vecX}, & \max_{\vecX} &\frob{ \calV, Q_\vecX}.
\end{align*}
Because \(\calM\) and \(\calV\) differ only by a diagonal matrix, and because the autocorrelation matrix of a random spin vector has \(\diag(Q_\vecX) = \One\), these problems are actually equivalent. 

The optimal random spin vector \(\vecX\) then corresponds to a distribution over optimal solutions to the quadratic integer programming problem
\begin{align*}
    \max_x x^*\mathcal{V}x \text{  st  } x\in \{\pm1\}^{\lvert V \rvert},
\end{align*}
which admits an interpretation as a form of the weighted Max Cut problem.

We can now use this form to define a multi-objective formulation for Hypergraph Partitioning. If we wish to move objects closer together according to a hypergraph \(H_1\) and move objects further apart according to a hypergraph \(H_2,\) we can utilize the imbalance matrix \(\calM_1\) of \(H_1\) and the variance matrix \(\calV_2\) of \(H_2\) to write
\begin{align*}
    \max_{\vecX}(\frob{\calM_1, Q_\vecX},\frob{ \calV_2, Q_\vecX}).
\end{align*}

The Pareto front of this optimization can then be estimated by evaluating the problem
\begin{align*}
    \max_{\vecX}&\langle \mathcal{O}, Q_\vecX\rangle_F & \mathcal{O} &\coloneqq \frac{\alpha}{\lambda_1} \calM_1 + \frac{(1-\alpha)}{\lambda_2} \calV_2
\end{align*}
where \(\lambda_1\) is the largest eigenvalue of \(\calM_1\), \(\lambda_2\) is the largest eigenvalue of \(\calV_2\), and \(\alpha\) is a series of values taken from \([0, 1]\). Including this scaling essentially compresses the spectrum of each matrix to \([0, 1],\) making it so that one term does not strictly dominate the other for a vast majority of tested alphas.

\section{Methodology}\label{sec:method}

In this section, we will introduce quantum frameworks for the Greatest Expected Imbalance/Least Expected Variance, Total Variance, and Multiobjective Hypergraph Partitioning problems. Each problem utilizes a shared QAOA ansatz construction based on a clique expansion of the input hypergraph. Furthermore, we define classical approximation algorithms for these problems based on hyperplane rounding, as well as an exact formulation. 

To develop our framework, we first present a unified formulation of the introduced problems. In essence, these problems can be described as finding the minimum autocorrelation matrix for a random spin vector \(\vecX\) which minimizes the greatest quadratic form given a collection of symmetric positive semidefinite matrices \(M_i\) for \(i \in \Index.\)
\begin{align}\label{eq:minimax_quad}
    \min_{\vecX}\max_{i \in \Index}\langle M_i, Q_\vecX\rangle_F
\end{align}
\subsection{Quantum Solvers}

Let \(\vecX\) be the random spin vector representing possible measurements along the computational (\(Z\)) basis from a quantum state \(\ket{\psi}\), that is \(\Prob[\vecX = x] = \abs{\braket{x}{\psi}}^{2}\). Note that for any bitstring $b$, we can define similar spin $x$ where $x_{i} = (-1)^{b_{i}}$. Then the autocorrelation of \(\vecX\) is given by
\begin{align*}
    (Q_\vecX)_{uv} &=\sum_x \Prob[\vecX = x] x_ux_v \\
    &= \sum_x \braket{\psi}{x}x_ux_v\braket{x}{\psi} =  \bra{\psi}Z_uZ_v\ket{\psi}
\end{align*}
since \(\ket{\psi} = \sum_x \braket{x}{\psi}\ket{x}\), \(\braket{x}{y} = 0\) for \(x\neq y \in \{\pm 1\}^{\abs{V}}\),  and \(Z_u \ket{x} = x_u\ket{x}\) for measurements along the computational ($Z$) basis\cite{nielsen2010quantum}.

% $Z_{u}\ket{x} = x\ket{x}$ as ($Z_{u}\ket{b} = (-1)^{b_{u}}\ket{b}$). Therefore, \(\bra{x}Z_uZ_v\ket{y} = x_u x_v\) if \(x = y\) and 0 otherwise for \(x,y \in \{\pm 1\}^{\abs{V}}\) for measurements along the computational ($Z$) basis \cite{nielsen2010quantum}.

Therefore, for a variational family of quantum states  $\ket{\psi(\theta)}$ this yields the following quantum variant of our problem
\begin{align*}
    \min_{\vecX, \theta}\max_{i \in \Index}\langle M_i, Q_\vecX\rangle_F \text{ st } (Q_\vecX)_{uv} = \bra{\psi(\theta)}  Z_uZ_v\ket{\psi(\theta)}
\end{align*}

Here, we tailor our ansatz construction to the well-known QAOA ansatz from \cite{farhi2014quantum} as introduced in \cref{sec:background}. To obtain a better approximation to our defined problem, we will use the multi-angle QAOA ansatz with low-depth \cite{herrman2022multi}. Given the hypergraph $H$, we construct the Mixer Unitary $U_{M}$ as usual where \[
U_{M}^{(l)}(\beta) = \prod_{u \in V} \exp{-i \beta_{u,l}X_{u}}.
\]

For the Cost Unitary, we include the term \(Z_uZ_v\) for each \(u \neq v\) such that there exists \(i \in \Index\) with \((M_i)_{uv} \neq 0\). For each of the problems described in this paper, this corresponds to taking the unweighted clique expansion of the input hypergraph $H$ (\cref{sec:hypergraph_partitioning}) as the cost graph. More specifically, for all \(u \neq v\) such that there exists \(i \in \Index\) with \((M_i)_{uv} \neq 0,\) then $U^{(l)}_{c}(\gamma) = \prod_{u, v} \exp{-i \theta_{(u, v),l}Z_u Z_v} $. A good thing when using this ansatz is that we can borrow the known theory from the literature \cite{kazi2025analyzing, shaydulin2021exploiting} to capture the dynamics of this ansatz, for example, as shown in \cite{bach2026learning}, where this ansatz can capture all cut distributions of given graph $G$.

To optimize the parameters of the circuit, we utilize gradient descent on a smoothed version of the objective; specifically, we replace the maximum used in \cref{eq:minimax_quad} with the LogSumExp function \cite{bach2026learning}.  This provides a smooth surface for the optimizer to work over. More details about this LogSumExp approximation can be found in \cite{bach2026learning}.

\subsection{Classical Solvers}

In this section, we present a pair of classical solvers which can be used to tackle the presented partitioning formulations. The first is an approximation algorithm based on hyperplane rounding following the Goemans-Williamson approach \cite{goemans1995improved}, while the second is an exact linear programming (LP) formulation which requires an exponentially large constraint matrix.

If we relax the spsd autocorrelation matrix of \cref{eq:minimax_quad} to an arbitrary spsd matrix, we get
\begin{align*}
    \min_{A\text{ spsd}}\max_{i \in \Index} \langle M_i, A\rangle_F \text{ st } \diag(A) = \One.
\end{align*}
which is equivalent to the semidefinite programming problem 
\begin{align*}
    \min_{A \text{ spsd}}\qquad & t\\
    \text{st}\qquad & \forall i \in \Index : \langle M_i, A\rangle \leq t\\
    &\diag(A) = \One.
\end{align*}
Since the above is convex, one can use an off-the-shelf solver like Clarabel to find an optimal solution to the relaxation in polynomial time, then use a singular value decomposition to acquire the factors \(A = BB^*\) \cite{clarabel}.

From here, we can follow the procedure of Goemans-Williamson and define a random spin vector via hyperplane rounding \(\vecX = \sign(B\mathbf{Y})\) where \(\mathbf{Y} \sim \mathrm{Normal}(0, 1).\)
As we introduced earlier, for any \(u,v\) we have 
\begin{align*}
    \frac{1}{4}\Exp\left[\abs{\vecX_u - \vecX_v}^2\right] =  \frac{1}{\pi}\arccos{(A_{uv})}
\end{align*}
Because of the inverse trigonometric identity \(\arccos(x) = \frac{\pi}{2} - \arcsin(x)\), we then have
\begin{align*}
    (Q_\vecX)_{uv} = \Exp[X_uX_v] = \frac{2}{\pi}\arcsin{(A_{uv})}.
\end{align*}
This provides a convenient way to compute the autocorrelation matrix of a hyperplane rounding distribution when it comes time to evaluate the objective.

It's worth noting, we can restrict our optimization problem only to random variables distributed by hyperplane rounding which yields
\begin{align*}
    \min_{\vecX, A\text{ spsd}}\max_{i \in \Index}\langle M_i, Q_\vecX\rangle_F \text{ st } (Q_\vecX)_{uv} = \arcsin(A_{uv}).
\end{align*}
Unfortunately, this formulation is itself NP-Hard due to the nonconvexity of \(\arcsin\) \cite{goemans1995improved}.

Additionally, we provide a linear programming algorithm for finding an exact solution. We do so by optimizing directly the underlying distribution \(q\) of \(\vecX\), which can be written as \(q_x = \Prob[\vecX = x]\). However, this requires the formation of an exponentially large matrix mapping each possible spin variable \(x \in \{\pm 1\}^{\lvert V \rvert}\) and index \(i \in \Index\) to the quadratic form \(C_{ix} \coloneqq x^*M_ix\). With this matrix, we can then define the linear program
\begin{align*}
    \min_q\qquad & t\\
    \text{st}\qquad & \forall i \in \Index : \langle C_i, q\rangle \leq t,\\
    & \langle \One, q\rangle = 1.
\end{align*}
As a result, whenever \(\lvert V \rvert\) remains relatively small, we have an effective way of finding the exact solution for the optimization problem, which we will utilize in order to form an effective test benchmark for the quantum algorithm.

\section{Analysis}\label{sec:analysis}

In this section, we will primarily discuss the relationship between hypergraph variance and other formulations of hypergraph partitioning. In particular, we will discuss how hypergraph variance is a direct generalization of traditional graph cut, and as such inherits any complexity results available for standard MAX CUT and balanced cut. Then, we will discuss how hypergraph variance relates to the standard objective for Hypergraph Partitioning.

Previously, we established that the quadratic form \(x^*\calV_e x\) can be written as the sum of squares 
\begin{align*}
    x^*\calV_e x &= \sum_{v} P_{ve}( x_v - p_e^*x)^2
\end{align*}
For a graph \(G\) where each edge has uniform edge contributions, that is \(P_{ve} = \frac{1}{\lvert e \rvert} = \frac{1}{2}\) whenever \(v \in e\),  then for each \(uv \in E\)
\begin{align*}
    x^*\calV_{uv} x & = \frac{1}{2}\left( \left( x_v - \frac{x_u + x_v}{2} \right)^2 + \left( x_u - \frac{x_u + x_v}{2} \right)^2\right) \\
    &= \frac{1}{4}( x_v - x_u)^2 = \frac{1 - x_ux_v}{2}.
\end{align*}
 Moreover, this implies that \(x^*\calV x = \frac{1}{4}x^*Lx\) where \(L\) is the Laplacian of \(G.\)  As a result, the Least Expected Variance problem is a direct generalization of Fair Cut Cover and the Total Variance problem is a direct generalization of Max Cut. In this way, we inherit a great deal of the complexity analysis available for these problems, as there is a direct reduction from these well known graph problems. For example, both problems are NP-Hard, as well as APX-Hard via this connection. Finally, this also implies that while assuming the Unique Games Conjecture, the proposed SDP approximation algorithm is optimal, as it would also serve as an approximation algorithm for Max Cut and Fair Cut Cover. % TODO maybe prove the equivalence more explicitly?

There also exists a relation from this problem to Hypergraph Partitioning and Set Splitting. Notably, 
for a vector \(x \in \{\pm 1\}^{\lvert V\rvert}\) the following statements are equivalent: (1) \(x\) does not cut \(e\), (2) \(x^*\calV_e x = 0\), and (3) \(\forall v \in e, p_e^*x = x_v\). This follows from basic properties of mean and variance: variance is zero if and only the underlying random variable is constant, and the mean of a collection of numbers on the boundary of a convex set is equal to one of those inputs if and only if it is equal to all of them.
Therefore we have
\begin{align*}
    \mathbbm{1}[x \text{ cuts } e] = \delta(x^*\calV_e x) = \sum_v P_{ve}\delta(x_v - m_e).
\end{align*}
Because \(x^*\calV_e x\) takes values in \([0, 1],\) \(\delta(x^*\calV_e x)\) is equivalent to \(\lceil x^*\calV_e x \rceil,\) meaning that the objective to Hypergraph Partitioning can be recovered as 
\begin{align*}
    \sum_{e,v}w_eP_{ve}\lceil x_v - m_e\rceil
\end{align*}
and we have  \(x^* \calV x \leq \Loss(x).\) In this way, Hypergraph variance can be viewed as a middle ground between Graph Partitioning and Hypergraph Partitioning.

% TODO monotonicity

% \subsection{Convergence}

\section{Experiments}\label{sec:experiments}

% TODO finish cleaning experiments

In this section, we present a series of experiments meant to demonstrate the quality of solutions accessible to our quantum algorithm relative to what can be achieved by the optimal polynomial-time approximation algorithm. All experiments are run on the Gen3 AMD EPYC-7352 system with 1~TB of main memory and an A100 GPU with 80~GB RAM. Exact solutions are computed using the Gurobi solver, while SDP solutions are computed using the Clarabel library \cite{clarabel}. Source code for reproducibility is available at \url{https://github.com/cameton/QuantumHypergraphPartitioning}.

The real-world hypergraphs in our experiments are subgraphs of the congress-bills and email-Enron hypergraphs from the dataset available at \url{https://www.cs.cornell.edu/~arb/data/} \cite{Benson_2018}. These subgraphs are found using the KaHyPar hypergraph partitioning library, yielding induced subgraphs with up to 20 vertices. Additionally, one subgraph from each collection is chosen to compare the quality of the Pareto front identified by QAOA with that found by SDP. We augment these results with a set of experiments on synthetic inputs, some randomly generated and one specifically constructed to be difficult for SDP.

 \begin{figure*}[ht]
    \centering
    \includegraphics[width=\linewidth]{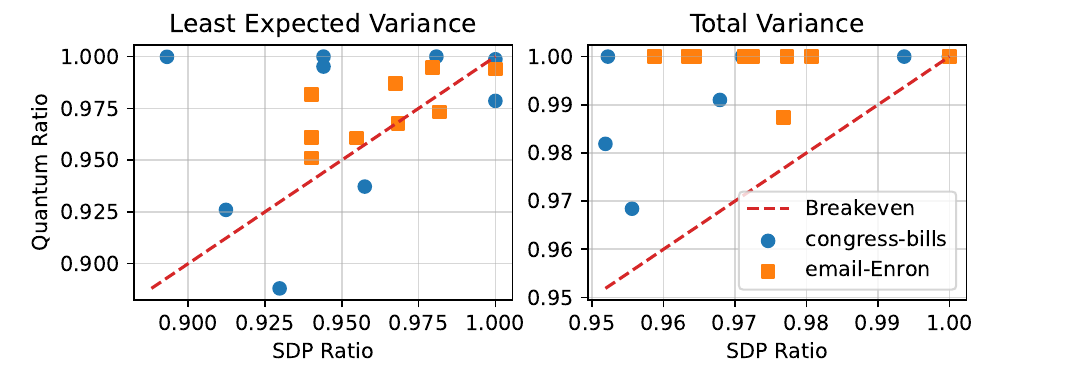}
    \caption{ % TODO Update Caption
    A comparison of the approximation ratios for QAOA and SDP on a collection of hypergraphs derived from the real world congress-bills (blue) and email-Enron (orange) hypergraphs from the Austin Benson hypergraph dataset for the Least Expected Variance and Total Variance problems. The red dotted line indicates the point at which both algorithms achieved an equivalent result, with dots above the line indicating that QAOA was favored. We see that for Least Expected Variance, QAOA was favored for all but 6 instances, while for Total Variance, QAOA was superior on every instance.}
    \label{fig:fair_results}
\end{figure*}

\subsection{Real World Data}

We compare the performance of the SDP approximation algorithm with our QAOA solver on 10 of the 50 subgraphs of congress-bills and all 8 of the subgraphs of email-Enron. For each hypergraph, we use a multi-angle QAOA ansatz with $p=3$ layers, trained for 300 iterations each with the stopping criterion after $30$ steps with accumulated improvement less than $10^{-4}$. In all of our experiments, we use the Adam optimizer with the following parameters (step size = $0.01$, $\beta_{1}=0.9$, $\beta_{2} = 0.99$, and $\epsilon = 1e-8$), where the gradient of the circuit is calculated using adjoint differentiation \cite{jones2020efficient}. For all experiments, we perform small-angle initializations where all initial angles are capped at $5e-2$.
We summarize the results of these experiments in \cref{fig:fair_results}.

% TODO update this once the results are rerun
Here we see that for the Total Variance problem, QAOA is superior to SDP for every tested instance. Notably, for a large fraction of tested instances, QAOA is able to achieve the optimal solution. For the Least Expected Variance problem however, SDP actually achieves a superior result on 6 of the tested instances. This makes some intuitive sense,  as we will later show in \cref{fig:layers_fair} and \cref{fig:layers_max} that QAOA for Total Variance may require fewer layers than QAOA for Least Expected Variance.

\begin{figure*}
    \centering
    \includegraphics[width=1.0\linewidth]{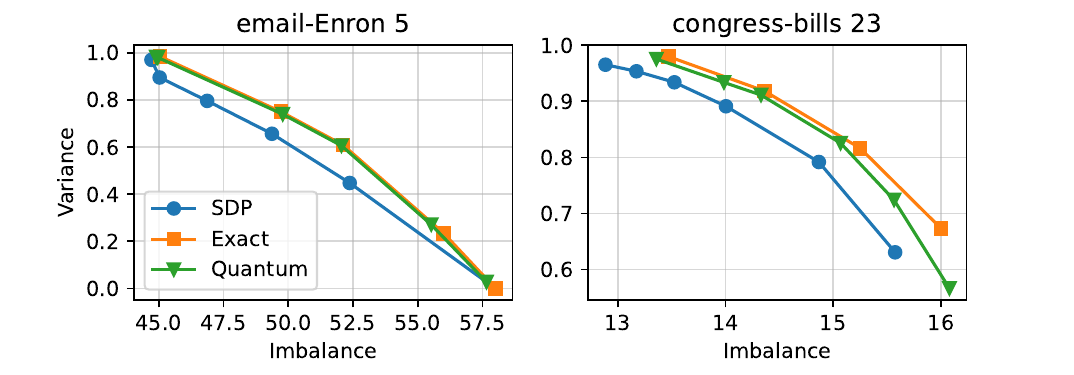}
    \caption{A depiction of the Pareto front achieved by quantum on the multiobjective optimization problem on the fifth hypergraph generated from the email-Enron dataset as well as the 23rd congress-bills subgraph. Each circle corresponds to a different \(\alpha\) used to evaluate that point in the Pareto front. We see for each instance a visible gap between SDP (blue) and quantum (green), which is often almost identical to the exact solution (orange) at each \(\alpha\) value.
    }
    \label{fig:multiobjective}
\end{figure*}

Additionally, we choose a single subgraph each for email-Enron and congress-bills on which to evaluate the Pareto front using the multiobjective partitioning formulation, which we show in \cref{fig:multiobjective}. For both instances, we evaluate the Pareto front for an evenly spaced grid of \(\alpha\)s form 0 to 1. When plotting these results, we show a portion of the Pareto curve in order to more clearly see the emerging gap between SDP and QAOA. For email-Enron, the points shown correspond to \(\alpha\)s between 0.5 and 0.75, while for congress-Bills we show \(\alpha\)s from 0.6 to 0.9. The Pareto front for QAOA hews closely to that found by the exact solver, and forms a curve around which dominates the front found by SDP.

\subsection{Synthetic Data}

Like graphs, hypergraphs admit a number of different models for generating random inputs \cite{random_hypergraph}. For example, for a fixed number of vertices \(n\) and number of edge \(M,\) one can generate a random \(n \times m\) incidence matrix \(B\) in an Erdos-Renyi style, where each entry \(B_{ve}\) is nonzero with probability \(p.\) For this paper, we use a convenient approximation based on the Poisson approximation of the binomial distribution. 

We define for each potential edge a random variable \(N_e \sim \mathrm{Poisson}(\mu - 2)\) for some rate \(\mu > 2.\) The size of \(e\) is then given by \(\min(\abs{V}, N_e + 2),\) and then we can then sample that many vertices without replacement to form the hyperedge. The weight of each hyperedge is then the number of times that unique set of vertices is sampled. We call these resulting graphs Poisson random hypergraphs. Conveniently, as long as \(\mu \ll \lvert V\rvert,\) the average size of each hyperedge is approximately \(\mu,\) so we can use this rate to control the average size of hyperedges.

\begin{figure*}
    \centering
    \includegraphics[width=\linewidth]{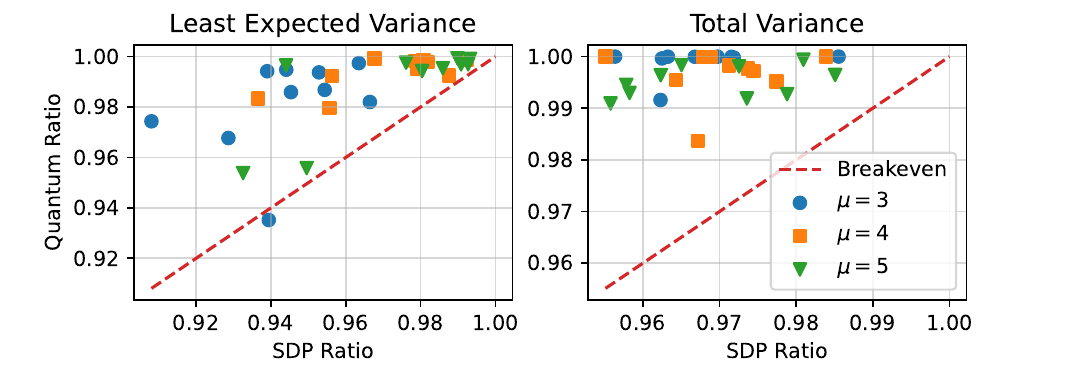}
    \caption{A comparison between Multi-angle QAOA and SDP approximation algorithm on a collection of Poisson random hypergraphs. Points above the breakeven line indicate that QAOA achieved a superior result, while points below the line indicate that SDP achieved a superior result. Inputs are grouped based on the average hyperedge size used during generation. Comparisons are made for the Least Expected Variance and Total Variance problems. For Least Expected Variance, QAOA is able to achieve a superior result for all but one instance, while for Total Variance, QAOA is superior on every instance.}
    \label{fig:poisson}
\end{figure*}

For our tests, we generate 30 Poisson random hypergraphs with 12 vertices and 16 edges; we generate 10 instances each for an average hyperedge size of 3, 4, and 5. We summarize the results for this experiment in \cref{fig:poisson}. Multi-angle QAOA is able to fairly conclusively triumph over SDP for this graph model, achieving a better result for the Least Expected Variance problem on all but one instance and achieving a better total variance on \textit{every} instance. Moreover, QAOA was able achieve the optimal solution for Total Variance on around 30\% of the tested instances. % TODO double check that number

\begin{figure}
    \centering
    \includegraphics[width=1.0\linewidth]{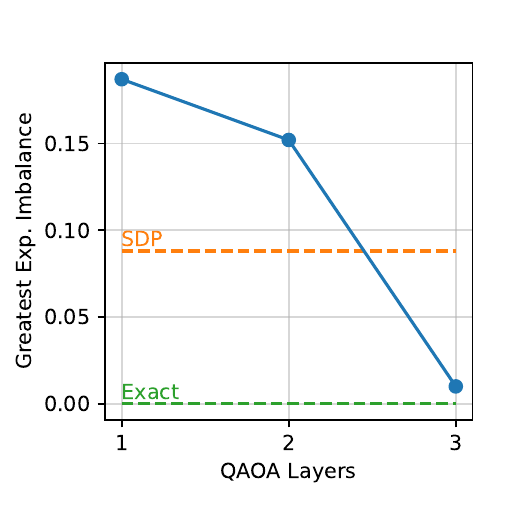}
    \caption{A demonstration of the improvement in solution quality achieved by increasing the number of layers in the QAOA ansatz. Results are provided for the Greatest Expected Imbalance problem on the hypergraph constructed from an enumeration of the cliques for the Karloff graph with parameters 6,3,1. For each, only 3 layers are required for QAOA to achieve a solution quality that exceeds that of SDP.}
    \label{fig:layers_fair}
\end{figure}

As another point of comparison, we construct a Hypergraph by enumerating all maximal cliques in one of the Karloff graphs, i.e., all maximal subsets of vertices where each pair of vertices is connected. These subsets form the hyperedges of the resulting hypergraph. Karloff \cite{karloff1996good} constructed these graphs to demonstrate that the $0.878$-approximation ratio of the Goemans-Williamson algorithm \cite{goemans1995improved} was tight, i.e., SDP struggles on these graphs. We specifically consider the Karloff graph with parameters $m=6, t=3, b=1$, which has a total of ${m \choose t} = {6 \choose 3} = 20$ vertices. We then look at how well the QAOA approach does for 1 to 3 layers, in order to show how solution quality increases. The clique enumeration of the Karloff graph is a highly symmetric structure, with 30 hyperedges all with size \(4.\) The optimal solution to the Greatest Expected Imbalance problem is actually $0$ for this graph. In \cref{fig:layers_fair}, we see that quantum is nearly able to achieve this result, getting an imbalance of only 0.01 with \(p = 3\) layers, exceeding SDP. 

\begin{figure}
    \centering
    \includegraphics[width=1.0\linewidth]{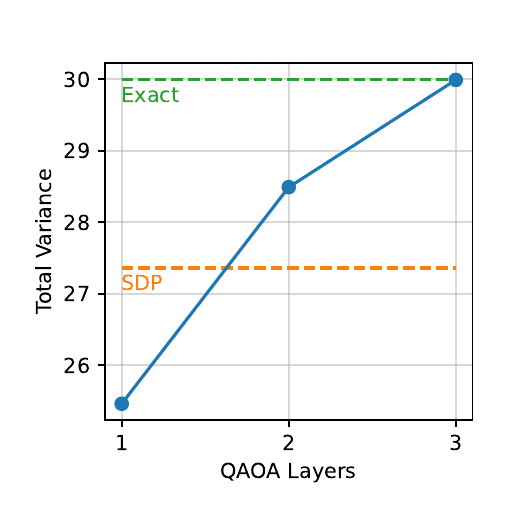}
    \caption{A demonstration of the improvement in solution quality achieved by increasing the number of layers in the QAOA ansatz. Results are provided for the Total Variance problem on the hypergraph constructed from an enumeration of the cliques for the Karloff graph with parameters 6,3,1. For each, only 3 layers are required for QAOA to achieve a solution quality that exceeds that of SDP.}
    \label{fig:layers_max}
\end{figure}

Similarly, we repeat this experiment on the Karloff graph for the total variance objective. In \cref{fig:layers_max} we see an even stronger result, that QAOA is able to exceed the solution of SDP with only two layers, again nearly achieving the optimum result. Both of these results are very promising, showing that it we can meaningfully compete with approachable classical methods with only a small number of layers.

\section{Discussion}\label{sec:discussion}

In this paper, we have presented a number of distributional formulations for the Hypergraph Partitioning problem to be solved with quantum computing. We present a QAOA-based algorithm for solving these problems, as well as an optimal polynomial-time approximation algorithm. Finally, we show experimentally that the QAOA approach is able to exceed the results of the polynomial-time algorithm with only 3 layers.

Importantly, the proposed distributional problems shift the role of quantum computation from merely searching for a single high-quality partition to learning a structured probability law over partitions. This distinction is important for hypergraph partitioning, where different hyperedges may impose competing requirements and no single partition need be uniformly favorable across all of them. A quantum state naturally encodes such trade-offs through its measurement distribution, allowing the optimizer to shape expected edge imbalance or variance directly. Consequently, the objective of distributional hypergraph partitioning is not just compatible with QAOA-style methods; it is also conceptually matched to them, since the algorithm’s output object is precisely the type of distribution that the problem asks us to optimize.

A natural next step for future work is to reformulate distributional hypergraph partitioning within the multilevel optimization paradigm and integration of quantum methods into it. Modern multilevel hypergraph partitioners \cite{schlag2023high,sybrandt2020hypergraph,shaydulin2019algdist,shaydulin2018aggregative} remain among the most scalable approaches in terms of quality/runtime trade-off, largely because they combine coarsening, coarse-level optimization, and refinement in a way that exposes structure at multiple resolutions. Integrating the proposed distributional and QAOA-inspired objectives into this framework would allow quantum or quantum-inspired solvers to be used selectively where they are most valuable, for example, on coarse hypergraphs, local refinement subproblems, or difficult boundary regions \cite{bach2024mlqaoa, ushijima2021multilevel}. This would also make it possible to study how probability distributions over partitions can be transferred across levels, replacing the standard deterministic projection of a single partition with a richer distributional prolongation and refinement process. Such a multilevel-distributional formulation could provide a practical bridge between the theoretical advantages of distributional optimization and the scalability requirements of real-world hypergraph partitioning.

\section*{Disclaimer}

The views expressed in this article are those of the authors and do not reflect the official policy or position of the U.S. Naval Academy, Department of the Navy, the Department of War, the U.S. Government, or Los Alamos National Laboratory.

\section*{Acknowledgments}
This work was supported in part by NSF award \#2444042. The research presented in this article was supported by the NNSA’s Advanced
Simulation and Computing Beyond Moore’s Law Program at Los Alamos National Laboratory. LANL report LA-UR-26-23383.

\bibliographystyle{unsrt}
\bibliography{qhgraph.bib,ilya-biblio}

\end{document}